\documentclass[11pt,a4paper]{amsart}

\usepackage[T1]{fontenc}
\usepackage[utf8]{inputenc}
\usepackage[english]{babel}
\usepackage{amsfonts, amssymb, amsmath, amsthm}
\usepackage[all]{xy}
\usepackage{enumerate}

\usepackage{graphicx}
\usepackage{psfrag}
\usepackage{geometry}
\usepackage[colorlinks=true,pdfstartview=FitV,linkcolor=blue,citecolor=green,urlcolor=black,filecolor=magenta]{hyperref}
\usepackage{verbatim}

\newtheorem{theorem}{Theorem}
\newtheorem{definition}[theorem]{Definition}

\newtheorem{cor}[theorem]{Corollary}

\newcommand{\C}{\ensuremath{\mathbb{C}}}
\newcommand{\N}{\ensuremath{\mathbb{N}}}
\newcommand{\Z}{\ensuremath{\mathbb{Z}}}

\newcommand{\R}{\ensuremath{\mathbb{R}}}
\newcommand{\B}{\mathcal B}

\newcommand{\Et}{\B_{top}}
\newcommand{\CL}{A_\Lambda}
\newcommand{\CLe}{A_\Lambda^{eigen}}
\newcommand{\T}{\ensuremath{\mathbb{T}}}
\newcommand{\TL}{\T_\Lambda}
\newcommand{\OL}{\Omega_\Lambda}

\newcommand{\vx}{{\bf x}}
\newcommand{\vk}{{\bf k}}
\newcommand{\vy}{{\bf y}}
\newcommand{\vo}{{\bf 0}}

\title{Topological Bragg peaks and how they characterise point sets}
\author{Johannes Kellendonk}
\address{Johannes Kellendonk\\ Universit\'e de Lyon,
Universit\'e Claude Bernard Lyon 1\\
Institute Camille Jordan, CNRS UMR 5208\\  69622 Villeurbanne, France}\email{kellendonk@math.univ-lyon1.fr}

\date{\today}

\begin{document}
\begin{abstract} Bragg peaks in point set diffraction show up as eigenvalues of a dynamical system. 
Topological Bragg peaks arrise from topological eigenvalues and determine the torus parametrisation
of the point set. We will discuss how qualitative properties of the torus parametrisation characterise
the point set.
  \end{abstract}
  
  \maketitle
  
\section{Introduction}
%It is a result of diffraction theory (in its Fraunhofer approximation) that 
The location $\vk$ of a %diffraction Bragg peak, that is 
Bragg peak in the $X$-ray diffraction picture of a material can be mathematically described as a point %in $\vk$-space 
for which $\hat\gamma(\{\vk\}) > 0$ \cite{Hof,Moody08}. 
Here $\hat\gamma$ is the Fourier transform of the autocorrelation of the material which is considered in an approximation in which the material
is modeled by a point set neglecting any kind of thermal fluctuations or other time evolution. 
%It turns out that, if one were only interested in the location of the Bragg peaks and not so much on their intensity, then 
The approach to point sets based on dynamical systems theory allows to give a more catchy way of saying which points may be the location of a Bragg peak: 
{\em $\vk$ may be the location of a Bragg peak if the plane wave $e^{i\vk\cdot \vx}$ with wave vector $\vk$ is in phase with the material.} 
It roughly means that the phase of the wave ought to be, up to a small error, the same at $\vx$ and at $\vy$ provided the local configurations around $\vx$ and $\vy$ are the same. A more precise formulation of this phrase needs a little effort and will be made below in a {\em topological} context
(Def.~\ref{def-top}). 
While diffraction theory is apriori not a topological theory, but rather of statistical nature,
%a measure theoretic one, 
%that is, claims are true in a statistical sense, 
this is for many structures including quasiperiodic ones not a short fall, because their diffraction is in some sense topological. 
Our intent is to show how this topological aspect of diffraction can be used to characterise point patterns. The results highlight that the concept of topological Bragg peaks is a fruitful one and it
would be interesting to find out whether they can be measured in an experiment.
%%%%%%%%%%%%%%%%%%%%%%%%%%%%%
\section{Topological Bragg peaks}
\subsection{Point patterns}
In this article we consider a particular class of point sets to which we simply refer to as  point patterns. 
Let $B(\vo,R)=\{\vy\in\R^n\,|\, \|\vx\|\leq R\}$ be the ball of radius $R$ centered at the origine and, for a point set $\Lambda$, by $\Lambda-\vx$ the point set shifted by $\vx$,
$\Lambda-\vx = \{\vy-\vx\,|\, \vy\in\Lambda\}$.
A {\em point pattern} 
$\Lambda\subset\R^n$ is a point set in $\R^n$ which satisfies the following conditions:
\begin{enumerate}
\item $\Lambda$ is {\em uniformly discrete}, i.e.\ there is a minimal distance between points.
%there exists $r>0$ such that any two distinct point have distance larger than $r$.
\item $\Lambda$ is {\em relatively dense}, i.e.\ there is $R>0$ so that any ball or radius $R$
contains a point of $\Lambda$. Points appear with bounded gaps. 
\item $\Lambda$ has {\em finite local complexity}, i.e., up to translation, one finds only finitely many local configurations of a given size. More precisely the collection of so-called $R$-patches
$\{B(\vo,R)\cap (\Lambda-\vx), \vx\in\Lambda\}$ is finite, and this for any choice of $R$,
\item  $\Lambda$ is {\em repetitive}, i.e.\ local configuration repeat inside $\Lambda$ 
with bounded gaps.
%in a relative dense way. More precisely the set 
%$\{\vy\in\R^n\,|\, B(\vo,R)\cap (\Lambda-\vx)=B(\vo,R)\cap (\Lambda-\vy)\}$ 
%is a relative dense set, and this for all choices of $\vx\in\Lambda$ and $R>0$. 
\end{enumerate}
Are these conditions realistic for describing atomic positions of materials? Condition~1 certainly is.
Condition~2 says that the material should not have arbitrarily large holes. Condition~3 is the strongest restriction and represents an idealisation which one can find in cut \& project sets used to describe ideal quasicrystals, but it would not allow for small random variations.
Having required Condition~3 the last condition seems a reasonable one to describe homogeneous materials. Let us add that from a mathematical point of view, Condition~3 is so far indispensible in order to obtain the kind of rigidity results we describe below.

Among the point patterns are Meyer sets and cut \& project sets.
%%%%%%%%%%%%%%%%
\subsubsection{Meyer sets}
%The concept of a Meyer set goes back to the work of
%Yves Meyer who introduced them in  [{\em Algebraic numbers and harmonic analysis}] 
%(the first under the name of harmonious sets) to study certain problems in number theory using Fourier analysis. 
A point set $\Lambda\subset\R^n$ is a {\em Meyer set} if it is relatively dense and the set of difference vectors $\Delta = \{\vx-\vy:\vx,\vy\in\Lambda\}$ is uniformly discrete. This is a very elementary geometric condition. Interestingly, the latter is equivalent to an analytic condition, namely that for all choices of $\epsilon>0$ the set
$\Lambda^\epsilon = \{ \vk\in\hat\R^n :  | e^{2\pi i \vk\cdot \vx} -1|\leq \epsilon, \forall 
\vx\in \Lambda\}$ is relatively dense. This says that the set of wave vectors for which the phase of the plane wave is, up to an error of $\epsilon$, equal to $1$ on all points of $\Lambda$, is relatively dense.
There are quite a few more equivalent conditions to the above (see \cite{Moody}) of which we mention one more: A set is a Meyer set if it is a relatively dense subset of a cut \& project set.

An example of a Meyer set which is not a cut \& project set can be derived from the famous
Thue-Morse substitution ${0}\mapsto {0}11{0},\: 1\mapsto 1{0}{0}1$. Iterating this substitution yields 
$${0}11{0} 1{0}{0}1 1{0}{0}1 {0}11{0}
1{0}{0}1 {0}11{0} {0}11{0} 1{0}{0}1 
1{0}{0}1 {0}11{0} {0}11{0} 1{0}{0}1 
{0}11{0} 1{0}{0}1 1{0}{0}1 0110 $$%100101100110$$%1001$$
which should be thought of as a finite part of a bi-infinite sequence. Now the subset 
$\Lambda\subset \Z$ given by the positions of the digit $1$ yields a Meyer set in $\R$, since
difference vectors are integer multiples of the unit vector in $\R$. 
%This is certainly uniformly discrete and hence $\Lambda$ is a Meyer set. 

Any Meyer set is uniformly discrete, relatively dense and has finite local complexity. So repetitive Meyer sets are point patterns.

\subsubsection{Cut \& project sets}
We assume that the reader is familiar with the concept of a cut \& project set, as it has been used since the early days of the discovery of quasicrystals for their description. 
%The same concept had appeared in the work of Yves Meyer \cite{Meyer72}
%under the name of model sets  and we will
We use the name {\em cut} \& {\em project set} synonymously for what is called  {\em model set} in the mathematics literature allowing the internal space of the construction to be more general than a vector space, namely to be a (locally compact) abelian group (see \cite{Moody}).
If the acceptance domain (or atomic surface) has a boundary whose measure is $0$ (this rules out many acceptance domains with fractal boundary) then the cut \& project set is called {\em regular}.
Any cut \& project set is a Meyer set. Ignoring a little somewhat ennoying detail 
we may say that a cut \& project set is repetitive.  

%%%%%%%%%%%%%%%%%%%%%%%%%%%%%%%%%%%%%%%%%%%%
\subsection{Pattern equivariant functions}
Given a point pattern $\Lambda\subset \R^n$ and a function $f:\R^n\to \C$ 
we want to make precise what it means for $f$ to depend only on the local configurations in the pattern. We have in mind a generalisation of the concept of a periodic function to which it specialised if $\Lambda$ were a periodic set. 
   
We say that $f:\R^n\to \C$ is {\em pattern equivariant}\footnote{In the literature once finds also the terminology {\em weakly pattern equivariant} for this, } for $\Lambda$
if for all $\epsilon>0$ there exists $R>0$ such that whenever the $R$-patches at $\vx$ and at $\vy$ are the same then $|f(\vx)-f(\vy)|<\epsilon$.  
Here we mean that   the $R$-patches at $\vx$ and at $\vy$ are the same if 
$B(\vo,R)\cap (\Lambda-\vx) = B(\vo,R)\cap (\Lambda-\vy)$, that is, the local configuration of size $R$ at $\vx$ is the same than the one at $\vy$ when they both have been shifted to the origin.
The example of a pattern equivariant function which the reader should have in mind is a potential energy function 
for a particle in a material whose atomic positions are given by $\Lambda$ each atom contributing to the potential energy with its local potential. 
%In this case $f(\vx) = \sum_{\vy\in\Lambda} v(\vx-\vy)$ where 
%$v:\R^n\to\C$ is a function which decays sufficiently fast for the sum to be well defined.

\begin{definition}\label{def-top} Let $\Et$ denote the set of 
vectors $\vk$ for which the plane wave $f_{\vk}(\vx)=e^{i\vk\cdot \vx}$ is pattern equivariant for $\Lambda$. 
\end{definition}
Thus $\vk\in\Et$ 
if the phase of the plane wave at a point $\vx$ is determined with more and more precision by the local configuration surrounding $\vx$; the larger the size of the configuration the more 
precise the phase is determined. 
Any $\vk\in\Et$ corresponds to a Bragg peak, although perhaps an extinct one, that is, a Bragg peak whose intensity is $0$. In this sense $\Et$ {\em is the set of locations of topological Bragg peaks for} $\Lambda$. Taking into account
extinct Bragg peaks may appear somewhat artifical but we gain the benefit that $\Et$ forms a group.  
Def.~\ref{def-top} does not involve a statistical ingredient but rests on continuity properties, which is   why we call the Bragg peak topological.
%It is most important that the locations of topological Bragg peaks form a subgroup of $\R^n$.
%
%Strictly speaking the above definition is a result of a theorem which is essentially based on Dworkin's argument. We will comment on that below.

\subsection{The dynamical system of a point pattern}
It is most useful to study the dynamical system associated with a point pattern. There are several versions of it which all more or less contain the same information. We present here the algebra version and the version based on a space: the continuous hull of $\Lambda$.

\subsubsection{Algebra version}
Consider the set $\CL$ of {\em continuous} functions $f:\R^n\to\C$ which are 
{\em pattern equivariant} for the point pattern $\Lambda$. $\CL$ is a commutative ($C^*$-) algebra under pointwise addition and multiplication. Moreover the group of translations $\R^n$ acts on $\CL$, that is, each vector of translation $\vx\in\R^n$ gives rise to a map $\alpha_{\vx}:\CL\to\CL$, namely
\begin{equation}\label{eq-action}
\alpha_{\vx}(f)(\vy) = f(\vy-\vx).
\end{equation}
This comes about as translation preserves the properties of a function to be continuous and pattern equivariant. The triple $(\CL,\R^n,\alpha)$ is called the (algebraic) dynamical system associated with 
$\Lambda$. The name "dynamical system" has nothing to do with a time evolution but  
is simply used by mathematicians for actions of groups (which in our case is $\R^n$, the group of space translations). 
\begin{definition}
An eigenvalue of the dynamical system  $(\CL,\R^n,\alpha)$ is a vector $\vk\in\R^n$ for which there exists a non-zero element $f\in\CL$ (its eigenfunction) such that 
\begin{equation}\label{eq-eigen}
 \alpha_{\vx}(f) = e^{2\pi i \vk\cdot\vx} f.
 \end{equation}
\end{definition}
It follows that $f$ must be a multiple of the plane wave, $f=cf_{\vk}$ with $c=f(\vo)$.
%$e^{2\pi i \vk\cdot\vx}$, 
%namely $f(\vx) = f(\vo)e^{2\pi i \vk\cdot\vx}$. 
Thus 
{\em a location of a topological Bragg peak is an eigenvalue of the dynamical system} $(\CL,\R^n,\alpha)$.

We now let $\CLe$ be the algebra generated by the eigenfunctions to eigenvalues of the system
$(\CL,\R^n,\alpha)$. The property of being an eigenfunction is preserved under translation and so we have a subsystem $(\CLe,\R^n,\alpha)$ of the system $(\CL,\R^n,\alpha)$. All what we will have to say depends on the relation between $\CLe$ and $\CL$.
%the subsystem and the main system.
\subsubsection{Torus parametrisation}
%Topological spaces and commutative $C^*$-algebras are dual to each other, namely 
To each commutative $C^*$-algebra corresponds a topological space in such a way that the algebra can be seen as the algebra of continuous functions over the space. This space is called the Gelfand spectrum of the algebra. The {\em continuous hull} $\OL$ of $\Lambda$ is the Gelfand spectrum of $\CL$, that is, $\CL\cong C(\OL)$. It has been subject to intensive study. Its elements are the point patterns which are locally indistinguishable from $\Lambda$, because they have the same local configurations. From a physical point of view, any element of $\OL$ is as good as $\Lambda$ to describe the material.

Now the action on $\CL$ becomes an action on $\OL$: $\alpha_{\vx}(\Lambda') = \Lambda'-\vx$. 
The triple $(\OL,\R^n,\alpha)$ is the space version of the dynamical system associated with 
$\Lambda$.

The Gelfand spectrum of $\CLe$ turns out to be a group $\TL$, in fact it is the (Pontrayagin) dual group of $\Et$. $\TL$ is a torus, or a limit of tori. 
The inclusion of $\CLe$ in $\CL$ as a sub algebra gives rise to a surjective map $\pi:\OL\to\TL$ which commutes with the actions. This map $\pi$ is called the {\em torus parametrisation}.
In the mathematical literature one calls $\TL$ also the maximal equicontinuous factor of $\OL$.
Our results below are based on the study of $\pi:\OL\to\TL$ and in particular, how close it is to a bijection.

\subsubsection{Topological conjugacy}
We say that two point patterns
$\Lambda$ and $\Lambda'$ are {\em topologically conjugate} if their associated dynamical systems are topologically conjugate, that is,
there exists a homeomorphism $\phi:\OL\to\Omega_{\Lambda'}$ which commutes with the actions.
If moreover $\phi(\Lambda)=\Lambda'$ then the topologically conjugacy is called pointed.

A well known example of a topological conjugacy is a local derivation which goes both ways, one says that $\Lambda$ and $\Lambda'$ are {\em mutually locally derivable} in this case. 
Topologically equivalent point patterns have the same dynamical properties, in particular they have the same locations of topological Bragg peaks and the same torus parametrization.
%%%%%%%%%%%%%%%%%%%%%%%%%%%%%%%%%
\subsubsection{Diffraction and the dynamical system}
We explain roughly how diffraction is related to the eigenvalues of the dynamical system. 
This goes back to Dworkin \cite{Dworkin} and was further developped (see, e.g.\ \cite{Moody08}). 
There is an ergodic probability measure $\mu$ on $\OL$ which has to do with the physical phase in which the material is and brings in the statistical aspect of diffraction. 
We may then look for solutions to (\ref{eq-eigen}) which have eigenfunctions which do not necessarily
belong to $\CL$, or equivalently to $C(\OL)$, but to the larger space $L^2(\OL,\mu)$ of functions on $\OL$ which are square integrable w.r.t.\ $\mu$. 
%Formula (\ref{eq-action}) can also be applied to such functions and so we get a definition of eigenvalues, exactly like in (\ref{eq-eigen}) exept that $f$ need merely be square integrable. 
We may therefore have more solutions and a larger group of eigenvalues $\B$. 
Let us call an eigenvalue an $L^2$-eigenvalue\footnote{the expression measurable eigenvalue is also used.} if it has an eigenfunction which is square integrable (but not necessarily continuous). Dworkin's arguement says that the location of any diffraction Bragg peak is an $L^2$-eigenvalue. Not every $L^2$-eigenvalue needs to come from a diffraction Bragg peak
%, because one may have extinction, 
%that is, a Bragg peak with intensity $0$. 
%But from the mathematical point of view it is better to work with eigenvalues, because they form a group.
but the group $\B$ is generated by the locations of Bragg peaks. The elements of 
$\B$ which do not come from a diffraction Bragg peak are interpreted as extinct (invisible) Bragg peaks.  The locations of topological Bragg peaks generate $\Et$ which is a subgroup of $\B$.
%are the eigenvalues with continuous eigenfunctions. They form a subgroup of all $L^2$-eigenvalues. A topological Bragg peak may correspond to a diffraction Bragg peak which is extinct and hence invisible. 
In this work, only $\Et$, that is, topological Bragg peaks play a role.
%%%%%%%%%%%%%%%%%%%%%%%%%%%%%%%%%%%%%%%%%%%
\section{Results}
We will present two kinds of results. For the first kind we assume that we have a repetitive Meyer set and obtain a characterisation depending on how close the torus parametrisation 
$\pi:\OL\to\TL$
is to a bijective map. For the second kind we assume only that we have a point pattern obtaining a partial classification of point patterns up to topological conjugacy. Recall that the torus parametrisation is always surjective. 
\subsection{Characterisation of repetitive Meyer sets}
\begin{theorem}
Let $\Lambda$ be a repetitive Meyer set. Then $\Et$ contains $n$ linear independent vectors  \cite{KS13}.
In other words, $\TL$ is at least as large as an $n$-torus $\T^n$. Furthermore
\begin{enumerate}
\item The torus parametrisation $\pi$ is injective on at least one point if and only if 
$\Lambda$ is a %repetitive 
{cut \& project set} \cite{Aujogue}.
\item The torus parametrisation $\pi$ is almost everywhere injective\footnote{this means that 
there exists a subset $\TL^0\subset\TL$ of measure $1$ 
such that each point of $\TL^0$ has a unique pre-image.} if and only if  $\Lambda$ is a %repetitive 
{regular} {cut \& project set} \cite{BLM06}.
\item $\pi$ is bijective if and only if $\Lambda$ is a periodic set (has $n$ independent periods)
\cite{BLM06,KellendonkLenz13}.
\end{enumerate}
\end{theorem}
\subsection{Classification of point patterns up to topological conjugacy}

\begin{theorem}[\cite{KS13}]
Let $\Lambda$ be a point pattern. $\Lambda$ is topologically conjugate
to a repetitive Meyer set if and only if
$\Et$ contains $n$ linear independent vectors.
\end{theorem}

\begin{cor} Let $\Lambda$ be a point pattern.

\begin{enumerate}
\item The torus parametrisation $\pi$ is injective on at least one point if and only if 
$\Lambda$ is a topologically conjugate to a %repetitive 
{cut \& project set}.
\item The torus parametrisation $\pi$ is almost everywhere injective
if and only if  $\Lambda$  is a topologically conjugate to a %repetitive 
{regular} {cut \& project set}.
\item $\pi$ is bijective if and only if $\Lambda$ is a periodic set (has $n$ independent periods).
\end{enumerate}
\end{cor}
\subsection{Beyond model sets}
In order to treat also cases in which there is no point on which
$\pi$ is injective we consider 
three numbers which measure how close $\OL$ sits above $\TL$. The first two are the maximal rank $Mr$, and the minimal rank $mr$, which are the largest, respectively smallest, number of elements the pre-image $\pi^{-1}(t)$ of $t$ can have when varying over $t\in\TL$. The really interesting third rank is the so-called coincidence rank. To define it we first introduce the relation that two elements
$\Lambda_1,\Lambda_2\in\OL$ are proximal ($\Lambda_1\sim_{prox}\Lambda_2$) if there exists a sequence $(\vx_k)_{k\in\N}\subset\R^n$ so that $\Lambda_1-\vx_k$ and $\Lambda_2-\vx_k$ coincide on a the patch of radius $k$ up to a translation of size smaller than $\frac1k$.
%$\lim_{k\to\infty} dist(\Lambda_1-x_k,\Lambda_2-x_k) = 0$.
This notion is more intuitive for Meyer sets: two Meyer sets $\Lambda_1,\Lambda_2\in\OL$
are proximal if and only if $\Lambda_1$ and $\Lambda_2$
agree on larger and larger balls. Now the {\em coincidence rank} $cr$ is defined to be the largest possible cardinality $m$ of a collection of elements  $ \Lambda_{1},\cdots,\Lambda_{m}\in\pi^{-1}({t})$ which satisfy  $\Lambda_i\not\sim_{prox}\Lambda_j$ ($i\neq j$). This number 
turns out not to depend on $t$.

Note that $ cr \leq mr\leq Mr$ and that the case $mr=1$ corresponds to cut \& project sets. 
Furthermore
$cr = mr$ whenever $\Omega_\Lambda$ contains an element which is not proximal to any other element. Primitive Meyer substitution tilings yield examples for which $cr=mr\leq Mr<\infty$
\cite{BargeKellendonk13}.
The Thue Morse substitution has $cr=2$.

\begin{theorem}
Let $\Lambda$ be a non-periodic point pattern. 
If {$cr$ is finite} then $\Lambda$ is topologically conjugate to a {Meyer set} and {$\Et\subset\R^n$ is dense}.
\end{theorem}

\subsection{How far does $(\Omega_\Lambda,\R^n)$ characterize $\Lambda$?}
The above classification of point patterns is up to topological conjugacy. We therefore need to understand to which extend topological conjugacy preserves the properties of a point set, like for instance the Meyer property. The first result in this direction is the following:
\begin{theorem}[\cite{KS13}]
Let $\Lambda\subset\R^n$ be a point pattern.
%\footnote{Repetitivity is actually not needed for this result.}.
$\Lambda$ and $\Lambda'$ are pointed  topologically conjugate whenever for all
$\epsilon>0$ exists a point pattern
$\Lambda_\epsilon$ such that $\Lambda$ and $\Lambda_\epsilon$ are {mutually locally derivable} %({block sliding code}) 
and $\Lambda_\epsilon$ and $\Lambda'$ are {shape conjugate}. Moreover,
within $\epsilon$ of each point of $\Lambda_\epsilon$ is a point of
$\Lambda'$ and vice versa. 
\end{theorem}
Here, a {\em shape conjugation} is a deformation of the pattern which preserves finite local complexity and induces a topological conjugacy. This notion
%is not very intuitive as it stands, but it may be very well 
may be formulated in the context of pattern equivariant cohomology \cite{CS2,K}. 
In fact, the shape conjugations of $\Lambda$ are classified by a subgroup of the first cohomology group of $\Lambda$. 
First investigations show that this group is rather small and so there are few shape conjugations.
%The question which properties of a point pattern are preserved by a shape conjugation is currently under investigation. 
Whereas shape conjugations of cut \& project sets with convex polyhedral acceptance domain
yield again cut \& project sets, we also found examples of more general cut \& project sets which allow for shape conjugations yielding point patterns which are not even Meyer sets \cite{KS}.


\begin{thebibliography}{99}
\thispagestyle{headings}
\markright{}

\bibitem{Aujogue}
J.-B.~Aujogue, PhD-thesis, Lyon, 2013.
%
%\bibitem[BSJ]{BJS}
%M.~Baake, M.~Schlottmann, and P.D. Jarvis, Quasiperiodic tilings with
%tenfold symmetry and equivalence with respect to local derivability, 
%{\em J. Phys. A} \textbf{24} (1991) 4637--4654.
  
%\bibitem[BM]{BM}
%M. Baake, R. V. Moody, Weighted Dirac combs with pure point diffraction, 
%{\em J.\ reine angew. Math.} {\bf 573} (2004) 61--94.
\bibitem{BLM06}
M. Baake, D. Lenz, R. V. Moody, Characterization of model sets by
dynamical systems,   
{\em Ergod. Th. \& Dynam. Systems} {\bf 26} (2006) 1-42.

\bibitem{BargeKellendonk13} M. Barge and J. Kellendonk,
Proximality and pure point spectrum for tiling dynamical systems, 
to appear in Michigan Math. Journal.


\bibitem{CS2} A.~Clark and L.~Sadun, 
When shape matters: deformations of tiling spaces, 
{\em Ergodic Theory Dynam. Systems} {\bf 26} (2006), 69--86. 

\bibitem{Dworkin} S.~Dworkin, Spectral theory and X-ray diffraction, 
{\em J. Math. Phys.} {\bf 34} (1993), 2964--2967.

%\bibitem[FS]{FS} N.P.~Frank and L.~Sadun, 
%Fusion: a general framework for hierarchical tilings of $R^d$,
%Preprint 2011, 	arXiv:1101.4930 


\bibitem{Hof} A.~Hof, On diffraction by aperiodic structures, 
{\em Commun. Math. Phys.} {\bf 169} (1995) 25--43.


\bibitem{K} J. Kellendonk, Pattern-equivariant functions and
cohomology, {\em J. Phys A.} {\bf 36} (2003), 5765--5772.


\bibitem{KellendonkLenz13} J. Kellendonk, D. Lenz, \textit{Equicontinuous Delone dynamical systems} , Canadian Journal of Mathematics \textbf{65} (2013), 149--170.

\bibitem{KS13} J. Kellendonk, L. Sadun, \textit{ Meyer sets, topological eigenvalues, and Cantor fiber bundles}, Journal of LMS, 2013.

\bibitem{KS} J. Kellendonk, L. Sadun, work in progress.

%\bibitem[La1]{Lagarias96} J.~Lagarias, Meyer's concept of quasicrystal and quasiregular sets. {\em Comm. Math. Phys.} {\bf 179} (1996) 365-376. 

%\bibitem[La2]{Lagarias} J.~Lagarias, Mathematical quasicrystals and 
%the problem of diffraction, in ``Directions in mathematical quasicrystals'',
%(M.~Baake and R.V.~Moody, eds.), CRM Monograph Series {\bf 13} (2000)
%61--93. 

%\bibitem{Meyer72} Y.~Meyer, Algebraic numbers and harmonic analysis. (1972), North-Holland. 

\bibitem{Moody} R.V.~Moody, Meyer sets and their duals, in
``The mathematics of long-range
aperiodic order'', (R.V.~Moody, ed.), Kluwer (1997), 403--441.

\bibitem{Moody08} R.V.~Moody, Recent developments in the mathematics of diffraction,
{\em Z.~Kristallogr.} {\bf 223} (2008) 795--800.
%
%\bibitem[St1]{St1} N.~Strungaru, 
%Almost periodic measures and long-range order in Meyer sets,
%{\em Discrete Comput. Geom.} {\bf 33} (2005), no. 3, 483--505. 

%\bibitem[St2]{St} N.~Strungaru, On the Bragg diffraction sprectra of a Meyer
%set, preprint 2010, arXiv:1003.3019.




\end{thebibliography}
\end{document}